\documentclass[10pt]{revtex4}
\usepackage{hyperref}
\usepackage{amsmath}
\usepackage[psamsfonts]{amssymb }
\usepackage{mathrsfs}

\begin{document}

\title{\Large Double Hodge Theory for a particle on Torus}

 \author{Vipul Kumar Pandey\footnote {e-mail address: vipulvaranasi@gmail.com}}
\author{ Bhabani Prasad Mandal\footnote {e-mail address: bhabani.mandal@gmail.com}}

\affiliation { Department of Physics, 
Banaras Hindu University, 
Varanasi-221005, INDIA.  }

\begin{abstract}
We investigate all possible nilpotent symmetries for a particle on torus. We explicitly construct four independent nilpotent BRST symmetries for such systems and derive the algebra between the generators of such symmetries. We show that such a system has rich mathematical properties and behaves as double Hodge theory. We further construct the finite field dependent BRST transformation for such systems by integrating the infinitesimal BRST transformation systematically. Such a finite transformation is useful in realizing the various theories with toric geometry.

\end{abstract}
\maketitle
\section{Introduction}
The formulation based on BRST symmetry \cite{1,2} plays crucial role in the discussion of quantization, renormalization, unitarity and other aspects of gauge theories. The nilpotency nature of BRST transformation is mainly responsible for simplified treatment in all these discussions. Thus it is extremely important to find more and more nilpotent symmetry associated with any system to study, particularly the systems with constraints. Toric geometry which is generalization of the projective identification that defines $CP^n$ corresponding to the most general linear sigma model provides a scheme for constructing Calabi-Yau manifolds and their mirrors \cite {3}. Recently, on the basis of boundary string field theory \cite {4}, the brane-antibrane system was exploited [5] in the toroidal background to investigate its thermodynamic properties associated with the Hagedorn temperature \cite {6,7}. The Nahm transform and moduli spaces of $CP^n$ models were also studied on the toric geometry \cite{8}. In a four dimensional, toroidally compactified heterotic string, the electrically charged BPS-saturated were shown to become massless along the hyper surfaces of enhanced gauge symmetry of a two-torus moduli subspace \cite {9}.\\ 
        \\ In the present work we investigate various possible nilpotent symmetries for a particle on torus. Usual BRST symmetry for a particle on torus has already been constructed \cite {10}. In this work we construct four different nilpotent symmetries associated with this system, namely BRST symmetry, anti-BRST symmetry, dual BRST (also known as co-BRST) symmetry and anti-dual BRST (also known as anti-co-BRST) symmetry \cite {11,12,13}. We further construct two different bosonic symmetries using these nilpotent BRST symmetries and some discrete symmetries associated with ghost number are also written for such systems. Complete algebra satisfied by charges, which generate these symmetries are derived. Deep mathematical connections of such system with Hodge theory \cite {33,34,35,36} are established in this work. We found that the system of particle on a torus is realized as Hodge theory w.r.t. to two different set of operators. The generators for BRST, dual-BRST symmetries and generator for corresponding bosonic symmetries constructed out of BRST and dual-BRST symmetries are analogous to exterior derivative, co-exterior derivative and Laplace operator in Hodge theory \cite {14,15,16,17,18,19,20,21,22}. On the other hand the charges corresponding to anti-BRST symmetry, anti-dual-BRST symmetry and bosonic symmetry constructed out of these two BRST symmetries also from set of de-Rham co-homological operators. This indicates the mathematical foundation of the theory of a particle on a torus is extremely rich.\\
        \\ We further extend the BRST transformation for this system by considering the BRST parameter as finite and field dependent. More than two decades ago Joglekar and Mandal introduced for the first time the concept of finite field dependent BRST(FFBRST)transformation \cite {37}, which had similar structure and properties of usual BRST transformation. However the path integral measure is not invariant due to finite nature of such transformation. It has been shown that by constructing suitable finite parameter one can calculate desirable Jacobian factor which under certain condition is added to the effective action of the theory . Thus FFBRST is capable of connecting generating functionals of two different effective theories. Because of these remarkable properties, FFBRST has become an useful tool of studying various field theoretic systems with BRST symmetry and it has found many applications \cite {38,39,40,41,42,43,44,45,46,47}. We have constructed FFBRST transformation for the system of particle on torus to show the connection between two theories on torus with different gauge fixing. Now we present the plan of this manuscript. 
        \\ We start with the brief introduction about the free particle on the the surface of torus in Sec. II.  Hamiltonian formulation for this theory is presented in sec. III. In Sec. IV the BFV formulation for this model has been discussed and BRST symmetry for such model has been constructed. In Sec. V the other nilpotent symmetry transformations for same system have been constructed. Co-BRST and anti co-BRST have been discussed in sec. VI. Other symmetries have been discussed in section VII. The connection between algebra satisfied by the nilpotent charges and de Rham co-homological operators of differential geometry is shown in Sec. VIII. In section IX we introduce FFBRST transformation and in next Sec. we connect theory in different gauges using FFBRST transformations. We conclude our results in Sec. XI.

\section{Free Particle on Surface of Torus}
A particle moving freely on the surface of a torus is described by Lagrangian
\begin{eqnarray}
L_0 = \frac{1}{2} m {\dot r}^2+\frac{1}{2} m r^2 {\dot \theta}^2+\frac{1}{2}m (b+r \sin\theta)^2{\dot\phi}^2
\end{eqnarray}
where $(r,\theta,\phi)$ are toroidal co-ordinates related to Cartesian coordinates as
\begin{equation}
x = (b+r\sin\theta)\cos\phi,
\quad y = (b+r\sin\theta)\sin\phi,
\quad z = r\cos\theta
\end{equation}

Here we have considered a torus with axial circle in the $x - y$ plane centered at the origin, of radius b, having a circular cross section of radius r. The angle $\theta$ ranges from 0 to $2\pi$, and the angle $\phi$ from 0 to $2\pi$. Since the particle moves on the surface of torus of radius r, it is constrained to satisfy  
\begin{equation}
\Omega_1 = r-a \approx 0
\end{equation}
The canonical Hamiltonian corresponding to the Lagrangian in Eq.(1) with the above constraint is then written as  
\begin{eqnarray}
H_0=\frac{p^2_r}{2m}+\frac{p^2_\theta}{2mr^2}+\frac{p^2_\phi}{2m(b+r\sin\theta)^2}+\lambda (r-a)
\end{eqnarray}
where $p_r$, $p_\theta$ and $p_\phi$ are the canonical momenta conjugate to the coordinate $r$, $\theta$ and $\phi$, respectively, given by
\begin{eqnarray}
p_r = m\dot r, \quad p_{\theta} = m r^2\dot\theta,\quad p_\phi = m (b+r\sin\theta)^2 
\end{eqnarray}
The time evolution of the constraint $\Omega_1$ yields the secondary constraint as
\begin{eqnarray}
\Omega_2 = p_r \approx 0
\end{eqnarray}

\section{Wess-Zumino term and Hamiltonian formulation} 
To construct a gauge invariant theory corresponding to the gauge non-invariant model in Eq.(4), we introduce the Wess-Zumino term \cite {23} in the Lagrangian density $\cal L$. For this purpose we enlarge the Hilbert space of the theory by introducing a new quantum field $\eta$, called as Wess-Zumino field, through the redifinition of fields r and $\lambda$ in the original Lagrangian density  $\cal L$ as follows
\begin{eqnarray}
r \rightarrow r-\eta; \quad    \lambda\rightarrow\lambda+\dot\eta
\end{eqnarray}
With this redefinition of the fields, the modified Lagrngian density becomes
\begin{eqnarray}
{\cal L^{I}} = \frac{1}{2} (\dot r - \dot \eta )^2+\frac{1}{2} m (r-\eta)^2{\dot\theta}^2	+ \frac{1}{2}m (b+(r-\eta)\sin\theta )^2{\dot\phi}^2-(\lambda+\dot\eta) (r-a-\eta)		  
\end{eqnarray}
 Canonical momenta corresponding to this modified Lagrangian density are then given by
 \begin{eqnarray}
p_r &=& m(\dot r - \dot \eta ),\quad p_{\eta} = (m(\dot r - \dot \eta )+(r-a-\eta)),\quad p_{\lambda} = 0\nonumber\\
p_\theta &=& m (r-\eta)^2{\dot\theta},\quad p_\phi = m (b+(r-\eta)\sin\theta )^2{\dot\phi}
 \end{eqnarray}
 The primary constraints for this extended theory is
\begin{equation}
\psi_1 \equiv p_{\lambda} \approx 0
\end{equation} 
The Hamiltonian density corresponding to $\cal L^I$ is written as,
\begin{equation}
H^{I} = p_r\dot r + p_{\eta}\dot\eta + p_{\theta}\dot\theta + p_{\phi}\dot\phi + p_{\lambda}\dot\lambda - \cal L^{I} 
\end{equation}
The total Hamiltonian density after the introduction of a Lagrange multiplier field u corresponding to the primary constraint $\psi_1$ is then obtained as 
\begin{equation}
H^{I}_T = \frac{p_r^2}{2m}+\frac{p^2_\theta}{2m(r-\eta)^2}+\frac{p^2_\phi}{2m(b+(r-\eta)\sin\theta)^2}+\lambda (p_r+p_\eta)+up_\lambda
\end{equation}
Following the Dirac's method of constraint analysis \cite{22,23,24,25}, we obtain secondary constraint
\begin{equation}
\psi_2 \equiv (p_\eta + p_r) \approx 0
\end{equation}

In next two sections, we extend this constrained theory to study the nilpotent symmetries associated with this theory. 
 
\section{BFV Formulation for free Particle on the Surface of Torus}
To discuss all possible nilpotent symmetries we further extend the theory using BFV formalism \cite{28,29,30,31}. In the BFV formulation associated with this system, we introduce a pair of canonically conjugate ghost fields (c,p) with ghost number 1 and -1 respectively, for the primary constraint $p_{\lambda} \approx 0$ and another pair of ghost fields $(\bar c,\bar p)$ with ghost number -1 and 1 respectively, for the secondary constraint, $(p_\eta + p_r) \approx 0$. The effective action for a particle on surface of the torus in extended phase space is then written as 
\begin{eqnarray}
S_{eff}& = &\int d^4 x \big[p_r\dot r + p_{\eta}\dot\eta + p_{\theta}\dot\theta + p_{\phi}\dot\phi - p_{\lambda}\dot\lambda
-\frac{p_r^2}{2m}-\frac{p^2_\theta}{2m(r-\eta)^2}\nonumber\\ & - &\frac{p^2_\phi}{2m(b+(r-\eta)\sin\theta)^2}+{\dot c} p + \dot{\bar c} \bar p - \{Q_b,\psi\} \big]
\end{eqnarray}
where $Q_b$ is the BRST charge and $\psi$ is the gauge fixed fermion. This effective action is invariant under BRST transformation generated by $Q_b$ which is constructed using constraints in the theory as
\begin{eqnarray}
Q_b = ic(p_r + p_{\eta})-i\bar p p_\lambda
\end{eqnarray}
The canonical brackets for all dynamical variables are written as  
\begin{eqnarray}
 [r, p_r] = [\theta, p_\theta] = [\phi, p_\phi] = [\eta, p_\eta] = [\lambda, p_\lambda] = \{\bar c,
\dot c\}= i, \{c, \dot{\bar c}\} = - i 
\end{eqnarray}
where rest of the brackets are zero. 
Now, the nilpotent BRST transformation, using the relation $s_b \phi =  [\phi, Q_b]_{\pm}$ ($\pm$ sign represents the fermionic and bosonic nature of the fields $\phi$), are explicitly written as
\begin{eqnarray}
s_b r &=& -c,  \quad  s_b\lambda = \bar p, \quad  s_b\bar p = 0, \quad s_b \theta = -c\nonumber\\
s_b p_\phi &=& 0, \quad s_b p_\theta = 0, \quad  s_b p = (p_r + p_\eta)\nonumber\\
s_b \bar c &=& p_\lambda, \quad s_b p_\lambda = 0, \quad s_b c = 0
\end{eqnarray}
In BFV formulation the generating functional is independent of gauge fixed fermion \cite{33}, hence we have liberty to choose it in the convenient form as
 \begin{eqnarray}
 \psi = p \lambda + \bar c (r + \eta + \frac{p_\lambda}{2})
 \end{eqnarray}
 Putting the value of $\psi$ in Eq. (14) and using Eqs., (15) and (16), we obtain
\begin{eqnarray}
S_{eff}& = &\int d^4 x \big[p_r\dot r + p_{\eta}\dot\eta + p_{\theta}\dot\theta + p_{\phi}\dot\phi - p_{\lambda}\dot\lambda
-\frac{p_r^2}{2m}-\frac{p^2_\theta}{2m(r-\eta)^2}\nonumber\\ 
& - &\frac{p^2_\phi}{2m(b+(r-\eta)\sin\theta)^2}+\dot c p + \dot\bar c \bar p + \lambda (p_r + p_\eta) + 2c\bar c \ - {\bar p} p + p_\lambda(r + \eta + \frac{p_\lambda}{2}) \big]   
\end{eqnarray}
and the generating functional for this effective theory is represented as
\begin{eqnarray}
Z_\psi &=& \int D \phi \quad  exp \big[iS_{eff} \big]  
\end{eqnarray}
Now integrating this generating functional over p and $\bar p$, we get 
\begin{eqnarray}
{Z_\psi} & = & \int D \phi' exp \big[i\int d^4 x \big[p_r\dot r + p_{\eta}\dot\eta + p_{\theta}\dot\theta + p_{\phi}\dot\phi - p_{\lambda }\dot\lambda-\frac{p_r^2}{2m}\nonumber\\ 
& - & \frac{p^2_\theta }{2m(r-\eta )^2} - \frac{p^2_\phi }{2m(b+(r-\eta )\sin\theta )^2}+\dot{\bar c} \dot c  + \lambda (p_r + p_\eta ) + 2c\bar c + p_\lambda (r + \eta + \frac{p_\lambda }{2}) \big]\big]    
\end{eqnarray}
where $D\phi'$ is the path integral measure for effective theory when integrations over fields p and $\bar p$ are carried out. Further integrating over field $p_\lambda$ we obtain an effective generating functional as    
\begin{eqnarray}
 Z_\psi &=& \int D \phi'' exp \big[i\int d^4 x \big[p_r\dot r + p_{\eta}\dot\eta + p_{\theta}\dot\theta + p_{\phi}\dot\phi -\frac{p_r^2}{2m}-\frac{p^2_\theta}{2m(r-\eta)^2}\nonumber\\ & - &\frac{p^2_\phi}{2m(b+(r-\eta)\sin\theta)^2}+\dot\bar c\dot c  + \lambda (p_r + p_\eta) - 2\bar c c - \frac{(\dot\lambda - r -\eta  )^2}{2} \big]\big]   
\end{eqnarray}
where $D\phi''$ is the path integral measure corresponding to all the dynamical variables involved in the effective action.The expression for effective action in above equation is similar to BRST invariant effective action in \cite{34}. The BRST symmetry transformation for this effective theory is written as  
\begin{eqnarray}
s_b r &=& - c, \quad s_b \lambda = \dot c,\quad s_b \eta = -c\nonumber\\
s_b p_r &=& 0 , \quad s_b p_\eta = 0\nonumber\\
s_b \bar c &=& - (\dot\lambda - \eta - r), \quad s_b c = 0
\end{eqnarray}
These transformations are on shell nilpotent.
\section{Nilpotent Symmetries}
In this section we will study various other nilpotent symmetries of this model with particle on a torus. For this purpose it is convenient to work using Nakanishi-Lautrup type auxiliary field B which linearize the gauge fixing part of the effective action in Eq.(22). The first order effective action is then given by
\begin{eqnarray}
S_{eff} &=& \int d^4 x \big[ p_r\dot r + p_{\eta}\dot\eta + p_{\theta}\dot\theta + p_{\phi}\dot\phi -\frac{p_r^2}{2m}-\frac{p^2_\theta}{2m(r-\eta)^2}\nonumber\\ & - &\frac{p^2_\phi}{2m(b+(r-\eta)\sin\theta)^2}+\dot{\bar c}\dot c  + \lambda (p_r + p_\eta) - 2\bar c c - B(\dot\lambda - r -\eta) + \frac {B^2}{2}\big ]  
\end{eqnarray}
 We can easily show that this action is invariant under the following off-shell nilpotent BRST transformation
\begin{eqnarray}
s_b r &=& - c,\quad s_b \lambda = \dot c,\quad s_b \eta = - c\nonumber\\
s_b p_r &=& 0, \quad s_b p_\eta = 0, \quad s_b \theta = 0\nonumber\\
s_b  \bar c &=& B, \quad s_b \bar c = 0, s_b p_\phi = 0\nonumber\\
s_b \phi &=& 0,\quad s_b p_\theta = 0
\end{eqnarray}
Corresponding anti-BRST transformation for this theory is then written by interchanging the role of ghost and anti-ghost field as
\begin{eqnarray}
s_{ab} r &=& - \bar c, \quad s_{ab} \lambda = \dot{\bar c}, \quad s_{ab} \eta = -\bar c\nonumber\\
  s_{ab} p_r &=& 0,  \quad s_{ab} p_\eta = 0,\quad s_{ab}p_\phi = 0\nonumber\\
s_{ab}  c &=&- B,  \quad s_{ab} \bar c = 0, \quad s_{ab} \theta = 0\nonumber\\
s_{ab} \phi &=& 0,   \quad s_{ab} p_\theta = 0   
\end{eqnarray}
The conserved BRST and anti-BRST charges $Q_b$ and $Q_{ab}$ which generate above BRST and anti-BRST transformations are written for this effective theory as
\begin{eqnarray}
Q_b = ic(p_r + p_{\eta})-i p_\lambda\dot c
\end{eqnarray}
and
\begin{eqnarray}
Q_{ab} = i\bar c(p_r + p_\eta) - ip_\lambda\dot{\bar c} 
\end{eqnarray}
Further by using following equation of motion 
\begin{eqnarray}
B + \dot p_r &=& 0,          \quad B + \dot p_\eta = 0\,             \quad  \dot r - p_r + \lambda = 0 \nonumber\\ 
\dot B &=& p_r + p_\eta , \quad \dot {\bar c} + 2\bar c = 0,               \nonumber\\ 
 {\dot c} + 2 c &=& 0,                   \quad B + \dot \lambda - r - \eta= 0 
\end{eqnarray}
it is shown that these charges are constants of motion i.e. $\dot Q_b = 0$, $\dot Q_{ab} = 0$, and satisfy following relations, 
\begin{eqnarray}
 Q_b Q_{ab} + Q_{ab}Q_b = 0
\end{eqnarray}
To arrive on these relations, the canonical brackets [Eq.(16)] of the fields and the definition of canonical momenta have been used
\begin{eqnarray}
p_{\lambda} = B, \quad p_{\bar c} = \dot c, \quad p_c = - \dot{\bar c} 
\end{eqnarray}
The physical states of theory are annihilated by the BRST and anti-BRST charges, leading to
\begin{eqnarray}
(p_r + p_\eta)|phys\rangle = 0
\end{eqnarray}
and
\begin{eqnarray}
p_\lambda|phys\rangle = 0 
\end{eqnarray}
This implies that the operator form of the first class constraint $p_\lambda \approx 0$ and $(p_r + p_\eta)\approx 0$ annihilates the physical state of the theory. Thus the physicality criteria is consistent with Dirac's method of quantization.
\section{Co-BRST and anti co-BRST symmetries}
 In this section, we investigate two other nilpotent transformations, namely co-BRST and anti co-BRST transformation which are also the symmetry of the effective action in Eq.(24). Further these transformations leave the gauge-fixing term of the action invariant independently and the kinetic energy term (which remains invariant under BRST and anti-BRST transformations) transforms under it to compensate for the transformation of the ghost terms. These transformations are also called as dual and anti dual-BRST transformation\cite{11,12,13}.

   The nilpotent co-BRST transformation $(s_d^2 = 0)$ and anti co-BRST transformation $(s_{ad}^2 = 0)$ which leave the effective action [in eq. (24)] for a particle on torus invariant, are given by
\begin{eqnarray}
s_d r &=& -\frac{1}{2}\dot{\bar c}, \quad s_d \lambda = - \bar c, \quad s_d \eta = -\frac{1}{2}\dot{\bar c}\nonumber\\
s_d p_r &=& 0 , , \quad s_d p_\eta = 0,\quad s_d \bar c = 0 \nonumber\\
s_d c &=& \frac{1}{2}(p_r + p_\eta), \quad s_d B = 0,  
\end{eqnarray} 
and  
\begin{eqnarray}
s_{ad} r &=& -\frac{1}{2}\dot c, \quad s_{ad} \lambda = - c, \quad s_{ad} \eta = -\frac{1}{2}\dot c\nonumber\\
s_{ad} p_r &=& 0 , \quad s_{ad} p_\eta = 0, \quad s_{ad} \bar c = 0 \nonumber\\
s_{ad} c &=& -\frac{1}{2}(p_r + p_\eta), \quad s_{ad} B = 0 
\end{eqnarray}
These transformations are absolutely anti-commuting as $\{S_d, S_{ad}\} = 0$.
The conserved charges for above symmetries are found using Noether's theorem and are written as 
\begin{eqnarray}
Q_d =  i\frac{1}{2}(p_r + p_\eta)\dot{\bar c} + i p_\lambda \bar c
\end{eqnarray}
and 
\begin{eqnarray}
Q_{ad} =  i\frac{1}{2}(p_r + p_\eta)\dot c + i p_\lambda c 
\end{eqnarray}
which generate the symmetry transformations in Eqs. (34) and (35) respectively. It is easy to verify the following relations
\begin{eqnarray}
s_d Q_d &=& - \{Q_d, Q_d\} = 0\nonumber\\
s_{ad} Q_{ad} &=& -\{Q_{ad}, Q_{ad}\}=0\nonumber\\
s_d Q_{ad} &=&-\{Q_{ad},Q_d\}=0\nonumber\\
s_{ad} Q_d &=&-\{Q_d,Q_{ad}\}=0
\end{eqnarray}
which reflect the nilpotency and anti-commutativity property of $s_d$ and $s_{ad}$ (i.e. $s_d^2 = 0$,$s_{ad}^2 = 0$ and $s_d s_{ad} + s_{ad} s_d = 0$).
\section{Other Symmetries}
In this section, we construct other symmetries related to this system. Two different bosonic symmetries are constructed out of four nilpotent symmetries. Discrete symmetry related to ghost number is also constructed.
\subsection{Bosonic Symmetry}
In this part we construct the bosonic symmetry out of these nilpotent BRST symmetries of the theory. The BRST $(s_b)$, anti-BRST $(s_{ab})$, co-BRST $(s_{d})$, and anti co-BRST$(s_{ad})$ symmetry operators satisfy the following algebra
\begin{eqnarray}
\{s_d, s_{ad}\} &=& 0,\quad \{s_b, s_{ab}\} = 0\nonumber\\
\{s_b, s_{ad}\} &=& 0,\quad \{s_d, s_{ab}\} = 0
\end{eqnarray}
and we define bosonic symmetries, $s_w$ and $s_{\bar w}$ as
\begin{eqnarray}
s_w\equiv\{s_b, s_d\},\quad s_{\bar w}\equiv \{s_{ab}, s_{ad}\} 
\end{eqnarray}
 The fields variables transform under bosonic symmetry $s_w$ as
\begin{eqnarray}
s_w r &=& -\frac {1}{2}(\dot B + p_r  + p_\eta),\quad   s_w \lambda = - \frac{1}{2}(2 B - \dot p_r - \dot p_\eta)\nonumber\\
s_w \eta &=& -\frac {1}{2}(\dot B + p_r  + p_\eta),\quad s_w p_r = 0, \quad s_w p_\eta = 0\nonumber\\
s_w c &=& 0,  \quad   s_w B = 0, \quad s_w \bar c = 0
\end{eqnarray} 
On the other hand transformation generated by $s_{\bar w}$ is
\begin{eqnarray}
 s_{\bar w} r &=& -\frac {1}{2}(\dot B + p_r  + p_\eta),\quad   s_{\bar w} \lambda =  \frac{1}{2}(2 B - \dot p_r - \dot p_\eta)\nonumber\\
s_{\bar w} \eta &=& -\frac {1}{2}(\dot B + p_r  + p_\eta),\quad s_{\bar w} p_r = 0,\quad s_{\bar w} p_\eta = 0\nonumber\\
s_{\bar w} c &=&0, \quad    s_{\bar w} B = 0, \quad s_{\bar w} \bar c = 0
\end{eqnarray}
However the transformation generated by $s_w$  and $s_{\bar w}$ are not independent as it is easy to see from Eq.(41) and (42) that the operators $s_w$ and $s_{\bar w}$ satisfy the relation $s_w + s_{\bar w} = 0$. This implies from Eq. (40), that 
\begin{eqnarray}
\{s_b, s_d\} = s_w = - \{s_{ab}, s_{ad}\}
\end{eqnarray}
It is clear from above algebra that the operator $s_w$  analogous of the Laplacian operator in the language of differential geometry and the conserved charge for the above symmetry transformation is calculated  as
\begin{eqnarray}
Q_w = - i [B^2 + \frac {1}{2} (p_r + p_\eta )^2]
\end{eqnarray}
which generates the transformation in Eq.(41).

Using equation of motion, it can readily be checked that 
\begin{eqnarray}
\frac {d Q_w} {dt} = - i \int dx [2B\dot B + (p_r + p_\eta )(\dot p_r + \dot p_\eta)] = 0 
\end{eqnarray}
Hence $Q_w$ is the constant of motion for this theory. 
\subsection{Ghost Symmetry and Discrete Symmetry}
Now we consider yet another kind of symmetry of this system called ghost symmetry. The ghost numbers of the ghost and anti-ghost fields are 1 and -1 respectively. Rest of the variables in the action of this theory have ghost number zero. Keeping this fact in mind we can introduce a scale transformation  of the ghost field, under which the effective action is invariant, as 
\begin{eqnarray}
c &\rightarrow &  e^\Lambda c\nonumber\\ 
\bar c &\rightarrow & e^{-\Lambda}\bar c\nonumber\\ 
\chi &\rightarrow & \chi
\end{eqnarray}
where $\chi = \{r,\eta,\theta,\phi,u,\lambda,p_r,p_\eta,p_\theta,p_\phi,p_u,B\}$ and $\Lambda$ is a global scale parameter. The infinitesimal version of the ghost scale transformation can be written as 
\begin{eqnarray}
s_g \chi &=& 0\nonumber\\
s_g c &=& c\nonumber\\ 
s_g \bar c &=& - \bar c 
\end{eqnarray}
The Noether's conserved charge for above symmetry transformation is calculated as
\begin{eqnarray}
Q_g = i[\dot {\bar c} c + \dot c \bar c]
\end{eqnarray}
In addition to above continuous symmetry transformation, the ghost sector respects the following discrete symmetry transformations
\begin{eqnarray}
c \rightarrow \pm i\bar c, \bar c \rightarrow \pm i c
\end{eqnarray} 
\section{Geometrical Cohomology}
In this section we study the de Rham cohomological operators and their realization in terms of conserved charges which generate the nilpotent symmetries for the theory of a particle on the surface of torus. In particular we point out the similarities between the algebra obeyed by de Rham co-homological operators and that by different BRST conserved charges.
\subsection{Hodge-de Rham decomposition theorem and differential operators}
Before we proceed to discuss the analogy, we briefly review the essential features of Hodge theory.
The de Rham cohomological operators in differential geometry obey the following algebra
\begin{eqnarray}
d^2 &=& \delta^2 =0, \quad \Delta = (d+\delta)^2 = d\delta + \delta d \equiv \{d,\delta\}\nonumber\\
\ [\Delta,\delta] &=& 0, \quad \ [\Delta, d] = 0
\end{eqnarray}
Where d, $\delta$ and $\Delta$ are exterior, co-exterior and Laplace-Beltrami operator respectively. The operator d and $\delta$ are adjoint or dual to each other and $\Delta$ is self-adjoint operator \cite{35}. It is well known that the exterior derivative raises the degree of form by one when it operates on forms $(i.e. df_n\sim f_{n+1})$, whereas the dual-exterior derivative lowers the degree of a form by one when it operates on forms $(i.e. \delta f_n\sim f_{n-1})$. However $\Delta$ does not change the degree of form $(i.e. \Delta f_n\sim f_n)$. $f_n$ denotes an arbitrary n-form object.
 
 Let M be a compact, orientable Riemannian manifold, then an inner product on the vector space $E^n(M)$ of n-forms on M can be defined as \cite{36}
\begin{eqnarray}
(\alpha, \beta)= \int_M \alpha\wedge* \beta
\end{eqnarray}
where  $\alpha,\beta\in E^n(M)$ and $*$ represents the Hodge duality operator \cite{37}. Suppose that $\alpha$ and $\beta$ are forms of degree n and $(n+1)$ respectively, the following relation for inner product will be satisfied
\begin{eqnarray}
(d\alpha,\beta) = (\alpha, \delta \beta)
\end{eqnarray}
Similarly, if $\beta$ is form of degree $n-1$, then we have the relation $(\alpha, d\beta) = (\delta\alpha,\beta)$. Thus the necessary and sufficient condition for $\alpha$ to be closed is that should be orthogonal to all co-exact forms of degree n. The form $\omega\in E^n(M)$ is called harmonic if $d\omega = 0$. Now let $\beta$ be a n-form on M and if there exists another n-form $\alpha$ such that $\Delta \alpha = \beta$, then for a harmonic form $\gamma\in H^n$,
\begin{eqnarray}
(\beta, \gamma) = (\Delta\alpha,\gamma)=(\alpha, \Delta\gamma)=0
\end{eqnarray}
where $H^n(M)$ denotes the subspace of $E^n(M)$ of harmonic forms on M. Therefore, if a form $\alpha$ exist with the property that $\Delta \alpha = \beta$, then Eq. (53) is necessary and sufficient condition for $\beta$ to be orthogonal to the subspace $H^n$. This reasoning leads to the idea that $E^n(M)$ can be partitioned into three distinct subspaces $\Lambda^n_d,\Lambda^n_\delta$ and $H^n$ which are consistent with exact, co-exact and harmonic forms respectively. Therefore, the Hodge-de Rham decomposition theorem can be stated as 

   A regular differential form of degree n$(\alpha)$ may be uniquely decomposed into a sum of the harmonic form $(\alpha)_H$, exact form $(\alpha_d)$ and co-exact form $(\alpha_\delta)$ i.e. 
\begin{eqnarray}
\alpha = \alpha_H + \alpha_d + \alpha_\delta
\end{eqnarray}
where $\alpha\in H^n$, $\alpha_s \in \Lambda^n_\delta$ and $\alpha_d \in \Lambda^n_d$
\subsection{Hodge-de Rham decomposition theorem and conserved charge}
The generators of all the nilpotent symmetry transformations satisfy the following algebra 
\begin{eqnarray}
Q_b^2 &=& 0, \quad Q_{ab}^2 = 0,\quad Q_d^2 = 0, \quad Q_{ad}^2 = 0 \nonumber\\
\{Q_b, Q_{ab}\} &=& 0,\quad \{Q_d, Q_{ad}\} = 0,\quad \{Q_b, Q_{ad}\} = 0 \nonumber\\
\{Q_d, Q_{ab}\} &=& 0,\quad [Q_g,Q_b] = Q_b,\quad [Q_g, Q_{ad}] = Q_{ad}\nonumber\\
\ [Q_g, Q_d] &=& -Q_d,\quad [Q_g, Q_{ab}] = - Q_{ab},\quad[Q_w, Q_r] = 0\nonumber\\
\{Q_b, Q_d\} &=& -\{Q_{ad}, Q_{ab}\} = Q_w
\end{eqnarray}
 Here the relations between the conserved charges $Q_b$ and $Q_{ad}$ as well as $Q_{ab}$ and $Q_{ad}$ can be found using equation of motions only. This algebra is similar to the algebra satisfied by de Rham co-homological operators of differential geometry given in Eq.(53). Comparing (53) and (58) we obtain following analogies
\begin{eqnarray}
(Q_b, Q_{ad})\rightarrow d, \quad (Q_d, Q_{ab})\rightarrow \delta,\quad Q_w\rightarrow \Delta 
\end{eqnarray}
 Let n be the ghost number associated with a given state $|\psi\rangle_n$ defined in the total Hilbert space of states, i.e.
\begin{eqnarray}
i Q_g|\psi\rangle_n = n |\psi\rangle_n
\end{eqnarray} 
Then it is easy to verify the following  relations
\begin{eqnarray}
Q_g Q_b|\psi\rangle_n &=& (n+1)Q_b|\psi\rangle_n\nonumber\\
Q_g Q_{ad}|\psi\rangle_n &=& (n+1)Q_{ad}|\psi\rangle_n\nonumber\\
Q_g Q_d|\psi\rangle_n &=& (n-1)Q_b|\psi\rangle_n\nonumber\\
Q_g Q_{ab}|\psi\rangle_n &=& (n-1)Q_{ad}|\psi\rangle_n\nonumber\\
Q_g Q_w|\psi\rangle_n &=& n Q_w|\psi\rangle_n
\end{eqnarray} 
which imply that the ghost numbers of the states $Q_b|\psi\rangle_n$,$Q_d|\psi\rangle_n$ and $Q_w|\psi\rangle_n$ are (n+1),(n-1) and n respectively. The states $Q_{ab}|\psi\rangle_n$ and $Q_{ad}|\psi\rangle_n$ have ghost numbers (n-1) and (n+1) respectively. The properties of set $(Q_b, Q_{ad})$ and $(Q_d, Q_{ab})$ are same as of operators d and $\delta$. It is evident from Eq. (58) that the set $Q_b, Q_{ad}$ raises the ghost number of a state by one and the set $Q_d, Q_{ab}$ lowers the ghost number of the same state by one. Keeping the analogy between charges of different nilpotent symmetries and Hodge-de Rham differential operators, we express any arbitrary state  $|\psi\rangle_n$ in terms of the sets $(Q_b, Q_d, Q_w)$ and $(Q_{ad},Q_{ab},Q_{\bar w})$ as 
\begin{eqnarray}
|\psi\rangle_n &=& |w\rangle_n + Q_b|\chi\rangle_{(n-1)} + Q_d|\phi\rangle_{(n+1)}\nonumber\\
\psi\rangle_n &=& |w\rangle_n + Q_{ad}|\chi\rangle_{(n-1)} + Q_{ab}|\phi\rangle_{(n+1)}
\end{eqnarray}
where the most symmetric state is the harmonic state$|w\rangle_n$ that satisfies 
\begin{eqnarray}
Q_w|w\rangle_n &=& 0, \quad Q_b |w\rangle_n = 0,\quad Q_d|w\rangle_n = 0\nonumber\\
Q_{ab}|w\rangle_n &=& 0, \quad Q_{ad}|w\rangle_n = 0
\end{eqnarray} 
analogous to the Eq. (53). Therefore the BRST charges for a particle on a torus forms two separate set of de-Rham co-homological operator, namely $\{Q_b, Q_{ab}, Q_w\}$ and $\{Q_d, Q_{ad}, Q_{\bar w}\}$.Thus we call the theory of a particle on torus as double Hodge theory. Fermionic charges $Q_b, Q_{ab}, Q_d$ and $Q_{ad}$ follow following physicality criteria
\begin{eqnarray}
Q_b|phys\rangle &=& 0, \quad Q_{ab}|phys\rangle = 0\nonumber\\
Q_d|phys\rangle &=& 0, \quad Q_{ad}|phys\rangle = 0
\end{eqnarray}
which lead to
\begin{eqnarray}
p_\lambda|phys\rangle &=& 0\nonumber\\
(P_r + P_\eta)|phys\rangle &=& 0
\end{eqnarray}
 This is the operator form of the first class constraint which annihilates the physical state as a consequence of physical criteria, which further is consistent with the Dirac's method of quantization of a system with first class constraints.
\section{Finite Field BRST transformations}
In this section we show that these nilpotent symmetries can be generalized by making the parameter finite and field dependent following the work of Joglekar and Mandal \cite{38}. The BRST transformations can be generated from BRST charge using relation $\delta \phi = [\phi, Q]\delta\Lambda$ where $\delta\Lambda$ is infinitesimal anti-commuting BRST parameter under which effective action remains invariant. Joglekar and Mandal generalized the anti-commuting BRST parameter $\delta\Lambda$ to be finite-field dependent instead of infinitesimal but space time independent parameter $\Theta[\phi]$. Under this generalization the path integral measure varies non-trivially. The Jacobian for these transformations for certain $\Theta[\phi]$ can be calculated by following way.  
\begin{eqnarray}
D\phi &=& J(k)D\phi'(k)\nonumber\\
       &=& J(k+dk)D\phi'(k+dk)
\end{eqnarray}
Where $k$ is a numerical parameter whose value lies between 0 and 1 ($0<_k<_1$). Here all the fields are taken to be $k$ dependent. For a field $\phi (x,k)$, $\phi(x,0) = \phi(x)$ and $\phi(x,k = 1) = \phi'(x)$. 

The invariance of the $S_{eff}$ under $\phi(x,0)\rightarrow \phi(x,k)$ is a BRST transformation given by
\begin{eqnarray}
\phi(0) = \phi(k) - \delta_b \phi(k)\Theta[\phi, k]. 
\end{eqnarray}

J($k$) can be replaced by $e^{iS_1[\phi(k);k]}$ for a certain functional $S_1$ which can be determined in each individual case using following condition
\begin{eqnarray}
\int D\phi(k)\big[\frac{1}{J(k)}\frac{d J(k)}{d k}-i\frac{dS_1}{dk}\big] e^{i (S_1 + S_{eff})} = 0 
\end{eqnarray}
where $\frac{dS_1}{dk}$ is a total derative of $S_1$ with respect to $k$ in which dependence on $\phi(k)$ is also differentiated and the Jacobian can be expressed as $e^{iS_1}$ where $S_1$ is local functional of fields which satisfies the Eq.(63) where change in Jacobian is calculated as
\begin{eqnarray}
\frac{J(k)}{J(k+dk)} &=& \sum_{\phi}{\pm}\frac{\delta \phi(x,k+dk)}{\delta \phi(x,k)}\nonumber\\
                     &=& \frac{1}{J(k)}\frac{d J(k)}{d k} d k
\end{eqnarray}
$\pm$ sign for bosonic and fermionic fields $(\phi)$ respectively.
\section{FFBRST for free particle on surface of torus}
 The effective action for the free particle on surface of torus using BFV formulation is written in Eq.(19) and its BRST transformation is given by Eq.(23).In BRST transformation given by Eq.(23), $\delta\Lambda$ is global, infinitesimal and anti-commuting parameter. FFBRST transformation corresponding to this BRST transformation is written as
 \begin{eqnarray}
s_b r &=& c \Theta, \quad s_b \lambda = -\dot c\Theta , \quad s_b \eta = c\Theta\nonumber\\
s_b p_r &=& 0, \quad s_b p_\eta = 0, \quad s_b c = 0\nonumber\\
s_b \bar c &=& (\dot\lambda - \eta - r)\Theta
\end{eqnarray}
where $\Theta$ is finite field dependent, global and anti-commuting parameter. Under this transformation too, effective action is invariant. 

 Generating functional for this effective theory can be written as
\begin{eqnarray}
Z_\psi &=& \int D \Phi \quad exp [i\int d^4 x \big[p_r\dot r + p_{\eta}\dot\eta + p_{\theta}\dot\theta + p_{\phi}\dot\phi - p_{\lambda}\dot\lambda -\frac{p_r^2}{2m}-\frac{p^2_\theta}{2m(r-\eta)^2}\nonumber\\ & - &\frac{p^2_\phi}{2m(b+(r-\eta)\sin\theta)^2}+\dot c p + \dot{\bar c} \bar p + \lambda (p_r + p_\eta) + 2c\bar c  - {\bar p} p + p_\lambda(r + \eta + \frac{p_\lambda}{2}) \big] ]   
\end{eqnarray}
where,
\begin{eqnarray}
D\Phi = dr dp_r d\theta dp_\theta d\phi dp_\phi d\eta dp_\eta d\lambda dp_\lambda dp d\bar p dc d\bar c 
\end{eqnarray} 
where $D\Phi$ is the path integral measure integrated over total phase space. The finite BRST transformation given above leaves the effective action invariant but path integral measure in generating functional is not invariant under this transformation. It gives rise to a Jacobian in the extended phase space which can be calculated as 
\begin{eqnarray}
D\Phi &=& dr dp_r d\theta dp_\theta d\phi dp_\phi d\eta dp_\eta d\lambda dp_\lambda  dp d\bar p dc d\bar c \nonumber\\
& = & J(k)dr(k) dp_r(k) d\theta(k) dp_\theta(k) d\phi(k) dp_\phi(k) d\eta(k) dp_\eta(k) d\lambda(k) dp_\lambda(k) du(k) dp_u(k) dp(k) d\bar p(k) dc(k) d\bar c(k) \nonumber\\
& = & J(k+dk)dr(k+dk) dp_r(k+dk) d\theta(k+dk) dp_\theta(k+dk) d\phi(k+dk) dp_\phi(k+dk) d\eta(k+dk) dp_\eta(k+dk)\nonumber\\ && d\lambda(k+dk) dp_\lambda(k+dk)  dp(k+dk) d\bar p(k+dk) dc(k+dk) d\bar c(k+dk) 
\end{eqnarray} 
Writing it in compact form as
 \begin{eqnarray}
&=&\int d^4 x \quad\sum_{\psi} \big[\frac{\delta \Psi(x, k + d k)}{\delta \Psi(x, k)}\big]
\end{eqnarray}
Where $\Psi = (r, p_r, \theta, p_\theta, \phi, p_\phi,\eta, p_\eta,\lambda, p_\lambda, p, \bar p, c, \bar c)$.
Which can be written as
\begin{eqnarray}
&=& 1 + dk\int \big [c\frac{\delta \Theta' (x,k+dk)}{\delta r(x,k)}- \dot c \frac{\delta \Theta (x,k+dk)}{\delta \lambda (x,k)}+ c\frac{\delta \Theta (x,k+dk)}{\delta \eta (x,k)}\nonumber\\ && + (\dot\lambda - \eta - r)\frac{\delta \Theta (x,k+dk)}{\delta \bar c(x,k)}\nonumber\\
&=&\frac{J(k)}{J(k+dk)}\nonumber\\
&=& 1 - \frac{1}{J(k)}\frac{d J(k)}{d k} d k
\end{eqnarray}
Now we consider an example to illustrate the FFBRST formulation. For that purpose we construct finite BRST parameter $\Theta$ obtained 
\begin{eqnarray}
\Theta' = i \gamma\int d^4 y \quad\bar c(y,k) p_\lambda(y,k) 
\end{eqnarray}
through
\begin{eqnarray}
\Theta=\int\Theta'(k)\quad d k
\end{eqnarray}  
The Jacobian change is calculated 
\begin{eqnarray}
\frac{1}{J(k)}\frac{d J(k)}{d k} = i \gamma\int d^4 y  \quad {p_\lambda}^2
\end{eqnarray}
We make an ansatz for $S_1$ as,
\begin{eqnarray}
S_1 = i\int d^4 x \quad\xi_1(k) {p_\lambda}^2
\end{eqnarray}
Where $\xi_1(k)$ is a $k$ dependent arbitrary parameter.  
Now,
\begin{eqnarray}
\frac{dS_1}{dk} = i\int d^4 x \quad\xi'_1(k) {p_\lambda}^2
\end{eqnarray}
Using condition in Eq. (65), we will get $\xi_1(k) = \gamma k$.
Now the modified generating functional can be written as,
\begin{eqnarray}
Z &=& \int D\chi'(k)\quad e^{i (S_1 + S_{eff})}\nonumber\\
&=&\int D \phi' \quad exp \big[i\int d^4 x \big[p_r\dot r + p_{\eta}\dot\eta + p_{\theta}\dot\theta + p_{\phi}\dot\phi - p_{\lambda}\dot\lambda -\frac{p_r^2}{2m}-\frac{p^2_\theta}{2m(r-\eta)^2}\nonumber\\ & - &\frac{p^2_\phi}{2m(b+(r-\eta)\sin\theta)^2}+\dot c p + \dot\bar c \bar p + \lambda (p_r + p_\eta) + 2c\bar c  - {\bar p} p + p_\lambda(r + \eta) +(\frac{\lambda'}{2}+\gamma k)  {p_\lambda}^2\big]\big]   
\end{eqnarray} 
Here generating functional at $k=0$ is the theory for a free particle on a surface of torus with a gauge parameter $\lambda'$ and at $k=1$, the generating functional for same theory with a different gauge parameter $\lambda'' = \lambda' + 2 \gamma$.  These two effective theories with two different gauge parameters on the surface of a torus are related through the FFBRST transformation with finite parameter given in Eq. (73). FFBRST transformation is thus helpful in showing the gauge independence of physical quantities.
 
\section{Conclusions}
 BFV formulation is a very powerful technique for quantization of gauge systems with constraints. It plays important role in analyzing the constraints and symmetries of the system. We have used this technique to study all the symmetries of a free particle on the surface of torus. We have constructed nilpotent BRST, dual-BRST, anti-BRST and anti-dual BRST transformations for this system. Dual-BRST transformations are also the symmetry of effective action and leaves gauge fixing part of the effective action invariant. Interchanging the role of ghost and anti-ghost fields the anti-BRST and anti-dual BRST symmetry transformations are constructed. We have shown that the nilpotent BRST and anti daul-BRST charges are analogous to the exterior derivative operators as the ghost number of the state $|\psi\rangle_n$ on the total Hilbert space is increased by one when these charges operate on this state and algebra followed by these operators is same as the algebra obeyed by the de Rham co-homological operators. Similarly the dual BRST and anti-BRST charges are analogous to co-exterior derivative. The anti-commutators of BRST and dual BRST and anti-BRST and anti dual-BRST charges lead to bosonic symmetry. The corresponding charges are analogous to Laplacian operator. Further, this theory has another nilpotent symmetry called ghost symmetry under which the ghost term of the effective action is invariant. We further have shown that this theory behaves as double Hodge theory as the charges for BRST $(Q_b)$, dual BRST $(Q_d)$ and the charges for the bosonic symmetry generated out of these two symmetries $(Q_w)$ form the algebra for Hodge theory. On the other hand charges for anti-BRST $(Q_{ab})$, anti-dual BRST $(Q_{ad})$ and $Q_{\bar w}$, charge for bosonic symmetry generalized out of these nilpotent symmetries also satisfy the Hodge algebra. Thus a particle on the surface of the torus has very rich mathematical structure.

                         We further constructed the FFBRST transformation for this system. By constructing appropriate field dependent parameter we have explicitly shown that such generalized BRST transformations are capable of connecting different theories on torus. It will be interesting to construct finite version dual BRST transformations and study its consequences in studying system with constraints.
                         
                         One of us (VKP) acknowledges University Grant Commission(UGC), India for its financial assistance under CSIR-UGC JRF/SRF scheme.
 
\ Disclaimer: There is no conflict of interest regarding the publication of this paper.

\end{document}